# TU Comae Berenices: Blazhko RR Lyrae Star in a Potential Binary System

## Pierre de Ponthière
*15 Rue Pré Mathy, Lesve, Profondeville 5170, Belgium; pierredeponthiere@gmail.com*

## Franz-Josef (Josch) Hambsch
*12 Oude Bleken, Mol, 2400, Belgium*

## Kenneth Menzies
*318A Potter Road, Framingham, MA 01701*

## Richard Sabo
*2336 Trailcrest Drive, Bozeman, MT 59718*

**Abstract**   We present the results of a photometry campaign of TU Com performed over a five-year time span. The analysis showed that the possible Blazhko period of 75 days published by the *General Catalogue of Variable Stars* is not correct. We identified two Blazhko periods of 43.6 and 45.5 days. This finding is based on measurement of 124 light maxima. A spectral analysis of the complete light curve confirmed these two periods. Besides the Blazhko amplitude and phase modulations, another long term periodic phase variation has been identified. This long term periodic variation affects the times of maximum light only and can be attributed to a light-travel time effect due to orbital motion of a binary system. The orbital parameters have been estimated by a nonlinear least-square fit applied to the set of (O-C) values. The Levenberg-Marquart algorithm has been used to perform the nonlinear least-square fit. The tentative orbital parameters include an orbital period of 1676 days, a minimal semi-major axis of 1.55 AU, and a small eccentricity of 0.22. The orbital parameter estimation also used 33 (O-C) values obtained from the SWASP survey database. Spectroscopic radial velocity measurements are needed to confirm this binarity. If confirmed, TU Com would be the first Blazhko RR Lyrae star detected in a binary system.

## 1, Introduction

The star TU Comae Berenices (TU Com) is classified in the *General Catalogue of Variable Stars* (Samus *et al.* 2011) as an RR Lyrae (RRab) variable star with a period of 0.4618091 day and a possible Blazhko period of 75 days. This period of 75 days was derived by Ureche (1965) from photographic observations. Using Fourier analysis of previous observations including ROTSE data (Wozniak *et al.* 2004), Sódor and Jurcsik (2005) questioned this Blazhko modulation. McGrath (1975), who observed this star at the Maria Mitchell Observatory (Nantucket, Massachusetts), did not detect a secondary modulation with a period of 75 days but one with a period of approximately 40 days.

Our results are based on 23,577 observations gathered during 150 nights between January 13, 2009, and May 23, 2015. The specifications of telescopes and CCD cameras used in this project and the number of observations for each telescope are provided in Table 1.

The CCD images were dark- and flat-field corrected with MAXIMDL software (Diffraction Limited 2004), and aperture photometry was performed using LESVEPHOTOMETRY (de Ponthière 2010), a custom software which also evaluates the SNR and estimates magnitude errors. The comparison stars are given in Table 2. The comparison star coordinates and magnitudes in B and V bands were obtained from the UCAC4 catalog (Zacharias *et al.* 2012). All the observations have been reduced with C1 as the magnitude reference and C2 as the check star. The observations were performed with a V filter and are not transformed to the standard system. The photometric observations were uploaded by the authors to the AAVSO International Database (Kafka 2015) where they can be retrieved.

Table 1. Telescope and camera specifications, numbers of observations, and photometric mean uncertainties.

| Location | Observer | Telescope Type | Camera Type | Number of Observations | Mean Uncertainty (mag.) |
|---|---|---|---|---|---|
| Cloudcroft, New Mexico | Hambsch | F/6.3 Meade 0.30m | SBIG ST9XM | 17252 | 0.014 |
| Mol, Belgium | Hambsch | Celestron 0.30m | SBIG ST8XME | 700 | 0.032 |
| Framingham, Massachusetts | Menzies | F/8 Hyperion 0.32m | SBIG STL-6303 | 2696 | 0.022 |
| Bozeman, Montana | Sabo | F/6.8 PlaneWave 0.43m | SBIG STL-1001 | 906 | 0.019 |
| Cloudcroft, New Mexico | de Ponthière | F/6.3 Meade 0.30m | SBIG ST-7 | 1301 | 0.022 |
| Lesve, Belgium | de Ponthière | F/6.2 Meade 0.20m | SBIG ST-7 | 722 | 0.024 |

Table 2. TU Com comparison stars.

| GSC Identification | UCAC4 Identification | R.A. (2000) h m s | Dec. (2000) ° ' " | B | V | B–V | Reference/Check |
|---|---|---|---|---|---|---|---|
| 2527-162 | 606-048343 | 12 13 40.55 | +31 00 46.22 | 14.868 | 14.167 | 0.701 | C1 |
| 2527-073 | 605-049038 | 12 14 18.96 | +30 59 22.82 | 15.126 | 14.473 | 0.653 | C2 |

All the data with an uncertainty larger than 0.050 magnitude have been eliminated from the dataset. The observations were not limited to the time of maxima, as can be seen in the folded light curve presented in Figure 1. This light curve is folded on the pulsation period determined in the next section. The photometric uncertainties for each telescope and location are provided in Table 1.

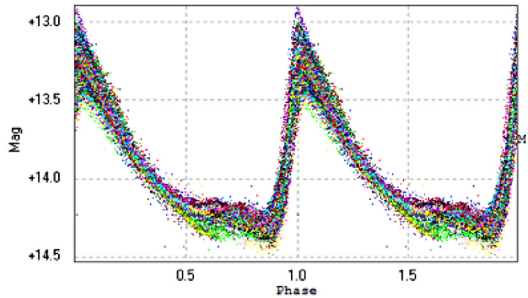

Figure 1. Folded light curve on the pulsation period.

## 2. Light curve maxima analysis

A custom software (de Ponthière 2010) fitting the light curve with a smoothing spline function (Reinsch 1967) was used to measure the times and magnitudes of light curve maxima. The observed times of light maxima are compared to a linear ephemeris to get the observed minus calculated (O–C) values. The (O–C) values and $M_{max}$ (Magnitude at Maximum brightness) of the 124 observed maxima are listed in Table 8 given in the Appendix.

A linear regression of (O–C) values has provided a pulsation period of 0.4618665 day, which has been used to establish the pulsation ephemeris.

$$HJD_{Pulsation} = (2456416.6221 \pm 0.0008) + (0.4618665 \pm 0.0000006) \, E_{Pulsation} \quad (1)$$

The origin of the ephemeris has been arbitrarily set to the highest recorded brightness maximum. The derived pulsation period is slightly different from the value of 0.4618091 published in the *General Catalogue of Variable Stars* (Samus *et al.* 2011). Figure 2 shows the (O–C) and $M_{max}$ values in the top and bottom panels, respectively.

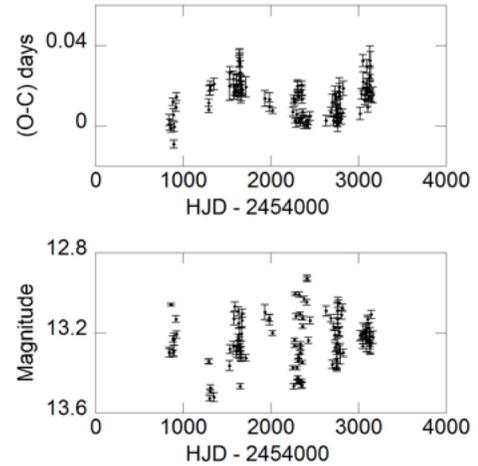

Figure 2. Top panel: O–C values, besides the Blazhko modulation of the times of maxima, a long term periodic variation is evident. Bottom panel: Magnitude at Maximum Brightness, the long term periodic variation seen in (O–C) is not apparent in the Magnitude at Maximum Brightness.

Besides the variations due to the Blazhko effect, a long term periodic variation of the times of maxima (O–C) is apparent from an inspection of Figure 2 (Top). This long term variation is not present in the magnitudes at maximum $M_{max}$ shown in Figure 2 (Bottom). The presence of this long term variation in the (O–C) and not in the $M_{max}$ can be explained by a light-travel time effect caused by an orbital motion around the common center of mass in a binary system.

RR Lyrae stars detected in binary systems are relatively rare; this is probably due to the technical challenges raised by the detection of the light-travel time effect in the observation datasets. At the end of the last century, TU UMa was the only one identified in a binary system (Saha and White 1990; Wade *et al.* 1999; Liska *et al.* 2015). Meanwhile, 12 RR Lyrae stars were recently discovered in the galactic bulge (Hajdu *et al.* 2015), and with data provided by the high precision photometry of the Kepler mission, other RR Lyrae stars in the galactic field have been identified as potential binary systems (Li and Qian 2014; Guggenberger and Steixner 2015). All those RR Lyrae stars are not affected by the Blazhko effect and do not show eclipses. The only RR Lyrae star detected in an eclipsing binary system is the non-classical RR Lyrae OGLE-BLG-RRLYR-02792, which has a mass of 0.26 $M_{\odot}$ (Pietrzynski *et al.* 2012).

To derive the (O–C) values, Hajdu *et al.* (2015) utilized the Hertzsprung (1919) method which compares the light curve to a template. This method is not appropriate for stars influenced by the Blazhko effect since their light curves do not repeat from one pulsation cycle to another. It is for this reason that Hajdu *et al.* (2015) eliminated stars impacted by the Blazhko effect from their investigations.

Table 3. TU Com frequency spectrum components obtained from light curve maxima.

| From | Frequency $(d^{-1})$ | $\sigma(d^{-1})$ | Period $(d)$ | $\sigma(d)$ | Amplitude | $\Phi$ $(cycle)$ | SNR |
|---|---|---|---|---|---|---|---|
| (O–C) values | 0.00061 | $1.3 \times 10^{-4}$ | 1634.8 | 356 | 0.00585 d | 0.198 | 19.1 |
| (O–C) values | 0.02209 | $0.78 \times 10^{-4}$ | 45.28 | 0.095 | 0.0229 d | 0.044 | 7.92 |
| $M_{max}$ values | 0.02306 | $2.2 \times 10^{-4}$ | 43.37 | 0.42 | 0.108 mag. | 0.056 | 14.8 |
| $M_{max}$ values | 0.02204 | $1.1 \times 10^{-4}$ | 45.36 | 0.22 | 0.082 mag. | 0.504 | 11.1 |

The periods and amplitudes of the (O–C) and $M_{max}$ values have been determined with PERIOD04 (Lenz and Breger 2005), a Fourier analysis and sine-wave fitting program. The results are presented in Table 3. Two Blazhko periods (43.37 and 45.36 d) are detected in the $M_{max}$ analysis but only one of those (45.28 d) is found in the (O–C) analysis. A long period (1634.8 d) is detected in the (O–C) analysis. As this long period is not detected in the $M_{max}$ analysis, it can be attributed to an orbital motion around a center of mass.

The two close Blazhko periods found in the magnitude at maximum spectrum are also detected in the spectral analysis of the light curve as shown in the next section. The presence of a main Blazhko period and another periodic modulation close to it has been reported for XZ Cyg by LaCluyzé, A., *et al.* (2004). They also detected long term variations of the main Blazhko period over a time span of several decades. Their analyses of XZ Cyg are based on observations covering a time span of several decades, which is not the case for our observations.

In order to detect a potential Blazhko period variation, we have created seasonal subsets of the magnitude at maximum values and applied a Fourier analysis followed by a sine-wave fitting. The results are presented in Table 4. The number of observations for the 2010 and 2012 seasons is too limited to perform a Fourier analysis and the corresponding subsets do not appear in Table 4.

It is unclear if the period variations are due to real Blazhko period deviation or to a non-repetitive Blazhko effect from one cycle to another.

TU Com was also observed by the robotic SuperWasp-North telescope (Butters *et al.* 2010) located on the island of La Palma (Spain) between 2004 and 2008. The star is identified as J121346.95+305907.6 in the SuperWASP database. From the light curves available on the SuperWASP website, 37 brightness maxima have been identified. Their measured (O–C) values are reported in the Appendix (Table 9). The magnitudes at maximum brightness are not reported in this table since it was not possible to reliably determine the offset between the SWASP magnitudes and the reference magnitudes used in our image reduction. The four maxima recorded in 2004 (JD 2453130 to 2453174) have large (O–C) values greater than 2 hours. These (O–C) values should be questioned and are not used in this paper; it is possible that an error occurred in the WASP automatic image reduction or in the data distribution process.

Table 4. TU Com period variation obtained from magnitude at maximum values.

| Subset $(year)$ | Period $(d)$ | $\sigma(d)$ | $N_{obs}$ |
|---|---|---|---|
| 2009 | 45.61 | 1.39 | 9 |
| 2011 | 43.90 | 0.82 | 25 |
| 2013 | 44.09 | 0.13 | 29 |
| 2014 | 44.49 | 0.27 | 26 |
| 2015 | 41.33 | 0.83 | 24 |

### 3. Frequency spectrum analysis of the light curve

The light curve of a Blazhko star may be considered as a signal modulated in amplitude and phase. The signal spectrum is characterized by a pattern of multiplets ($kf_o \pm nf_B$) based on a pulsation frequency $f_o$ and Blazhko modulation frequency $f_B$. Generally, from ground-based observations, only the central triplets are detected, as the other components are hidden in the noise. The amplitudes, phases, and uncertainties of the spectral components have been obtained with PERIOD04 by performing successive Fourier analyses, pre-whitenings, and sine-wave fittings. Only the components having a signal to noise ratio (SNR) greater than 3 have been retained as significant signals.

Table 5 provides the complete list of spectral components. Besides the pulsation frequency $f_o$ and its harmonics $nf_o$, two groups of triplets corresponding to Blazhko periods and a component based on the suspected orbital period have been found. The frequencies and periods corresponding to Blazhko frequencies $f_{B1}$ and $f_{B2}$ and to an orbital period are given in Table 6. The two Blazhko periods corresponding to $f_{B1}$ and $f_{B2}$ are close to the periods found in the analysis of magnitude at maximum brightness. The orbital period of 1,601 days is in relatively good agreement with the value of 1,634.8 days found in the (O–C) analysis.

During the sine-wave fitting, the pulsation frequency $f_o$, ($f_o - f_{B1}$), ($2f_o + f_{B2}$), and ($2f_o - f_{Orb}$) have been left unconstrained and the other frequencies have been forced as combinations of the four unconstrained frequencies. The uncertainties of frequencies, amplitudes, and phases estimated from Monte Carlo simulations have been multiplied by a factor of two as it is known that the Monte Carlo simulations underestimate these uncertainties.

Figure 3 presents the (O–C) values pre-whitened with the assumed orbital period of 1,601 days versus time. By comparison with the top panel of the Figure 2, it can be seen that the long term variation is effectively removed and only variations due to the short term Blazhko effect remain.

The same (O–C) pre-whitened data folded with the 43.66-

Table 5. TU Com multi-frequency fit results.

| Component | $f$(d$^{-1}$) | $\sigma(f)$ | $A_i$ | $\sigma(A_i)$ (mag.) | $\Phi_i$ | $\sigma(\Phi_i)$ (cycle) | SNR |
|---|---|---|---|---|---|---|---|
| $f_o$ | 2.165128 | $8.69 \times 10{-7}$ | 0.3998 | 0.0017 | 0.3296 | 0.0006 | 117.1 |
| $2f_o$ | 4.330256 | | 0.2050 | 0.0017 | 0.9899 | 0.0015 | 60.7 |
| $3f_o$ | 6.495383 | | 0.1178 | 0.0017 | 0.7559 | 0.0021 | 32.3 |
| $4f_o$ | 8.660511 | | 0.0636 | 0.0018 | 0.4953 | 0.0042 | 16.8 |
| $5f_o$ | 10.82564 | | 0.0400 | 0.0016 | 0.2296 | 0.0075 | 11.3 |
| $6f_o$ | 12.99077 | | 0.0302 | 0.0018 | 0.9825 | 0.0080 | 9.7 |
| $2f_o - f_{Orb}$ | 4.329631 | $8.75 \times 10{-6}$ | 0.0530 | 0.0016 | 0.7854 | 0.0059 | 15.7 |
| $f_o - f_{B1}$ | 2.142221 | $11.1 \times 10{-6}$ | 0.0300 | 0.0016 | 0.7745 | 0.0086 | 9.1 |
| $f_o + f_{B1}$ | 2.188034 | | 0.0395 | 0.0019 | 0.0218 | 0.0084 | 10.3 |
| $3f_o - f_{B1}$ | 6.472477 | | 0.0171 | 0.0015 | 0.1028 | 0.0086 | 4.6 |
| $3f_o + f_{B1}$ | 6.51829 | | 0.0294 | 0.0015 | 0.4460 | 0.0143 | 8.1 |
| $2f_o + f_{B2}$ | 4.352244 | $12.9 \times 10{-6}$ | 0.0367 | 0.0018 | 0.8150 | 0.0075 | 10.0 |
| $f_o - f_{B2}$ | 2.143139 | | 0.0344 | 0.0019 | 0.4456 | 0.0326 | 10.1 |
| $f_o + f_{B2}$ | 2.187117 | | 0.0100 | 0.0017 | 0.1479 | 0.0083 | 3.0 |

Table 6. TU Com triplet component frequencies and periods.

| Component | Derived from (d$^{-1}$) | Frequency (d$^{-1}$) (d) | $\sigma$ (d) | Period (d) | $\sigma$ |
|---|---|---|---|---|---|
| $f_o$ | | 2.165128 | $8.69 \times 10{-7}$ | 0.461867 | $1.85 \times 10{-7}$ |
| $f_{B1}$ | $f_o - f_{B1}$ | 0.022906 | $1.11 \times 10{-5}$ | 43.66 | 0.02 |
| $f_{B2}$ | $2f_o - f_{B2}$ | 0.021989 | $1.30 \times 10{-5}$ | 45.48 | 0.02 |
| $f_{Orb}$ | $2f_o - f_{Orb}$ | 0.000625 | $8.9 \times 10{-6}$ | 1601.0 | 22.9 |

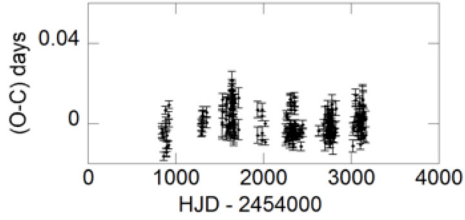

Figure 3. (O–C) values pre-whitened with the 1601-day assumed orbit period.

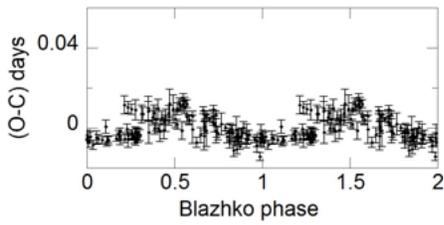

Figure 4. (O–C) values pre-whitened with the 1,601-day assumed orbit period and folded with the 43.66-day Blazhko period.

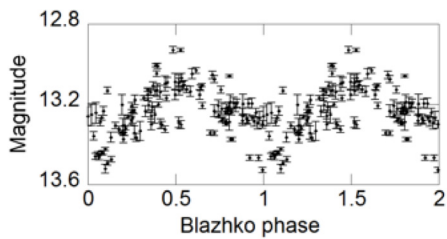

Figure 5. Magnitude at maximum brightness folded with the 43.66-day Blazhko period.

day Blazhko period are shown in the phase diagram of Figure 4, and the phase diagram of the magnitude at maximum using the same Blazhko period is given in Figure 5. In the phase diagrams of (O–C) values and magnitudes at maximum, the remaining scatter of the data is likely due to the presence of the second Blazhko period or to the non-repetitive Blazhko effect from one cycle to another.

## 4. Orbital parameter estimation

A pulsating star residing in a binary system can be seen as a regular "clock" in orbit around a center of mass. This orbital motion will affect the times of light maxima. For a pulsating star not affected by the Blazhko effect, besides a possible secular pulsation rate acceleration/deceleration, the orbital motion will be the only source of variations of the (O–C) values. Those (O–C) variations allow the evaluation of the orbital parameters by a non-linear least square fit with respect to the light-travel time equation. When applied to Blazhko pulsating stars, the Blazhko effect will be considered as noise affecting the (O–C) measurements. The Blazhko effect will increase the uncertainties of the orbital parameter estimation.

The light-travel time equation due to orbital motion is given by Hilditch (2001):

$$\tau = \frac{(a_{RRL} \sin i)}{c} \frac{(1 - e^2)}{(1 + e \cos \nu)} \sin (\nu + \omega) + \tau_0 \qquad (2)$$

where $a_{RRL}$ is the semi-major axis, $e$ is the eccentricity, $\nu$ is the true anomaly, $i$ is the orbit inclination, $\omega$ the periastron longitude, and $c$ the speed of light. Without the additional term $\tau_0$, the zero-point of $\tau$ is reached when the star is at the same distance as the mass center of the binary system, that is, when $\nu + \omega = \pm k \pi$. The zero-point of the (O–C) values obtained in section 2 has been arbitrarily set to the time of the highest recorded light maximum. The additional offset $\tau_0$ is introduced to compensate for the difference between these two zero-points.

The true anomaly can be calculated from:

$$\tan \frac{v}{2} = \sqrt{\frac{1+e}{1-e}} \tan \frac{E}{2} \qquad (3)$$

where E is the eccentric anomaly which is evaluated by solving the Kepler equation:

$$E - e \sin E = 2\pi \, (t - T_{peri}) / P_{orb} \qquad (4)$$

and where $T_{peri}$ is the epoch of periastron passage and $P_{orb}$ is the orbital period. The semi-major axis a and the orbital inclination i are linked without additional information on the secondary star.

To obtain the estimation of orbital parameters ($a_{RRL}$ sin i / c, e, ω, $P_{orb}$, $T_{peri}$, $\tau_0$) the Levenberg-Marquardt algorithm was used to minimize the sum of squares of the residuals $r_i = (O–C)_i - \tau \, (t_i, \, \boldsymbol{\beta})$ where $t_i$ are the observed times of maxima and $\boldsymbol{\beta}$ the vector of parameters ($a_{RRL}$ sin i / c, e, ω, $P_{orb}$, $T_{peri}$, $\tau_0$). For each observed time of maxima, the light-travel time $\tau \, (t_i, \, \boldsymbol{\beta})$ is obtained by solving the Kepler equation (4) and calculating Equations (3) and (2).

The orbital parameter estimation was performed with the 124 (O–C) values derived from our observations (Table 8) and with 33 (O–C) values obtained from the SuperWASP survey (Table 9). The four (O–C) values corresponding to maxima recorded in 2004 were eliminated as they are abnormally large and are in question.

The results of the least-square fit are:

$a_{RRL}$ sin i / c = 0.00893 d (1.55 AU)
$P_{orb}$ = 1,676 d = 4.59 years
e = 0.22
ω = −0.978 rad
$T_{peri}$ = 2455006 HJD
$\tau_0$ = 0.0117 d

Using these orbital parameters, the theoretical light-travel times have been calculated and are compared to the (O–C) values in Figure 6.

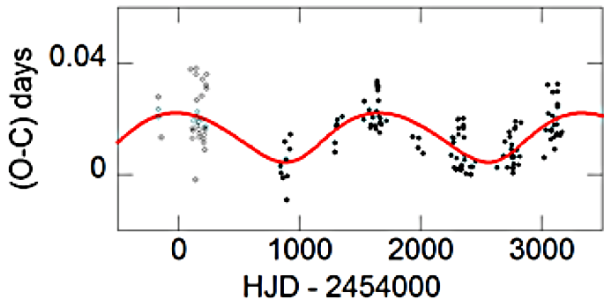

Figure 6. The (O–C) values (black diamonds) already presented in Figure 2 are compared to the light-travel time (red line) calculated from the orbital parameter solution. The (O–C) values derived from the SWASP database for the years 2005 and 2006 are represented with open diamonds.

The estimated orbital parameters are relatively uncertain as the scatter of the (O–C) values used for the orbital parameter estimation is as large as the orbit light-travel time variations.

An estimation of the semi-amplitude of the star's radial velocity may be derived from Equation (5):

$$K = (2\pi a_{RRL} \sin i) / (1 - e^2)^{1/2} = 10.3 \text{ km/s.} \qquad (5)$$

From Kepler's third law, the mass function for a barycentric orbit M = $(4\pi^2 \, a_{RRL}{}^3) / P_{orb}{}^2$ is related to star masses through M = $(G \, m_s{}^3) / (m_{RRL} + m_s)^2$, where $m_{RRL}$ and $m_s$ are the masses of TU Com and the secondary star, respectively, and G is the gravitational constant. If $m_{RRL}$ and $m_s$ are expressed in solar masses $M_\odot$ and $P_{orb}$ in years, G is equal to $4\pi^2$ and Kepler's third law can be rewritten as

$$\frac{m_s{}^3 \sin^3 i}{(m_{RRL} + m_s)^2} = \frac{a_{RRL}{}^3 \sin^3 i}{P_{orb}{}^2} \qquad (6)$$

Assuming a classical RR Lyrae mass of 0.7 $M_\odot$ for TU Com, the minimum mass of the secondary star ($m_s \sin i$) may be evaluated by solving Equation (6) rewritten as a third order polynomial. With the numerical values of 1.55 AU for (a sin i) and 4.59 year for $P_{orb}$ the real solution of the polynomial provides a minimal mass ($m_s \sin i$) for the secondary star fortuitously equal to 0.70 $M_\odot$. The third order polynomial has been solved for other orbital inclinations and the results are provided in Table 7.

Table 7. Secondary mass and semi-major axes of the two stars for different orbital inclinations.

| Orbital Inclination (degrees) | Secondary Mass ($M_\odot$) | $a_{RRL}$ (AU) | $a_s$ (AU) |
|---|---|---|---|
| 90 | 0.70 | 1.55 | 1.54 |
| 80 | 0.72 | 1.57 | 1.53 |
| 70 | 0.77 | 1.65 | 1.50 |
| 60 | 0.87 | 1.78 | 1.44 |
| 50 | 1.07 | 2.02 | 1.32 |
| 40 | 1.45 | 2.40 | 1.16 |
| 30 | 2.36 | 3.09 | 0.92 |
| 20 | 5.56 | 4.52 | 0.57 |
| 10 | 34.84 | 8.90 | 0.18 |

RR Lyrae stars are old stars and the secondary star probably formed at the same epoch and would have the same metallicity. With these assumptions, it may be assumed that the secondary star is in a more evolved state than the RR Lyrae star and eventually it ended as a white dwarf, because massive stars evolve more rapidly than lower mass ones. It is possible that the secondary star brightness is not large enough to allow spectroscopic measurement of its radial velocity.

If the radial velocities of the two stars may be measured, the mass ratio may be derived from the relationship:

$$\frac{m_s}{m_{RRL}} = \frac{V_{RRL}}{V_s} = \frac{a_{RRL}}{a_s} \qquad (7)$$

This relationship also shows that the chances to measure the secondary radial velocity are reduced when the mass is larger.

## 5. Conclusions

This observational campaign and data analysis has shown that the Blazhko period of 75 days mentioned in the *General Catalog of Variable Stars* is not correct. Alternatively, two Blazhko periods of 43.6 and 45.5 days have been identified from a light curve maxima analysis and confirmed from the spectral analysis of the light curve. The origin of the two Blazhko periods remains unclear; it could be due to a real second period or to a variation of a main Blazhko period or to the non-repetitive Blazhko effect from one cycle to another. A long term periodic variation of the (O–C) values suggests that TU Com is in a binary system with an orbital period of about 1,676 days. A tentative set of orbital parameters have been derived from a non-linear least square fit of the (O–C) values with respect to the light-travel time equation. The authors intend to continue their photometric observations in future years to extend the amount of data and to refine these results. They also invite other amateur astronomers to join their campaign. In order to confirm the binarity of TU Com, it is suggested that this star be integrated into a spectroscopic radial velocity measurement campaign like the study started by Guggenberger *et al.* (2015). Applying the radial velocity method to determine the orbital parameters will be a challenge as the radial velocities are also impacted by the pulsation/motion of the atmospheric layers which will be additionally affected by the Blazhko effect.

## 6. Acknowledgements

The AAVSO is acknowledged for the use of AAVSOnet telescopes at Cloudcroft (New Mexico). The authors thank Dr. K. Kolenberg and Prof. Dr. G. Rauw for their help with the secondary star evolution analysis and the referee for the comments which helped to clarify and improve the paper. This work has made use of data from DR1 of the WASP data (Butters *et al.* 2010) as provided by the WASP consortium, and the computing and storage facilities at the CERIT Scientific Cloud, reg. no. CZ.1.05/3.2.00/08.0144, which is operated by Masaryk University, Czech Republic.

**Appendix**

Table 8. TU Com measured brightness maxima.

| Maximum HJD | Error | O–C (day) | E | Magnitude | Error | Maximum HJD | Error | O–C (day) | E | Magnitude | Error |
|---|---|---|---|---|---|---|---|---|---|---|---|
| 2454844.8935 | 0.0028 | 0.0030 | −3403 | 13.297 | 0.025 | 2456352.9023 | 0.0027 | 0.0178 | −138 | 13.463 | 0.014 |
| 2454849.9716 | 0.0026 | 0.0006 | −3392 | 13.301 | 0.020 | 2456353.8284 | 0.0021 | 0.0201 | −136 | 13.441 | 0.010 |
| 2454861.9785 | 0.0020 | −0.0011 | −3366 | 13.059 | 0.007 | 2456358.9022 | 0.0021 | 0.0134 | −125 | 13.347 | 0.009 |
| 2454881.8518 | 0.0020 | 0.0120 | −3323 | 13.231 | 0.016 | 2456363.9763 | 0.0015 | 0.0070 | −114 | 13.168 | 0.011 |
| 2454888.7734 | 0.0029 | 0.0056 | −3308 | 13.302 | 0.016 | 2456364.8983 | 0.0013 | 0.0052 | −112 | 13.128 | 0.008 |
| 2454892.9157 | 0.0020 | −0.0089 | −3299 | 13.291 | 0.026 | 2456376.9021 | 0.0018 | 0.0005 | −86 | 13.032 | 0.015 |
| 2454901.6995 | 0.0024 | −0.0005 | −3280 | 13.232 | 0.030 | 2456410.6207 | 0.0017 | 0.0029 | −13 | 13.047 | 0.014 |
| 2454918.7984 | 0.0020 | 0.0093 | −3243 | 13.131 | 0.016 | 2456414.7748 | 0.0015 | 0.0002 | −4 | 12.929 | 0.017 |
| 2454924.8080 | 0.0019 | 0.0146 | −3230 | 13.209 | 0.018 | 2456416.6221 | 0.0009 | 0.0000 | 0 | 12.93 | 0.008 |
| 2455292.4472 | 0.0015 | 0.0081 | −2434 | 13.343 | 0.012 | 2456427.7097 | 0.0022 | 0.0028 | 24 | 13.238 | 0.015 |
| 2455293.3744 | 0.0018 | 0.0116 | −2432 | 13.344 | 0.012 | 2456452.6526 | 0.0022 | 0.0049 | 78 | 13.139 | 0.017 |
| 2455305.3889 | 0.0024 | 0.0175 | −2406 | 13.527 | 0.013 | 2456630.9308 | 0.0026 | 0.0027 | 464 | 13.091 | 0.026 |
| 2455310.4699 | 0.0025 | 0.0180 | −2395 | 13.491 | 0.012 | 2456687.7446 | 0.0024 | 0.0069 | 587 | 13.127 | 0.018 |
| 2455311.3954 | 0.0024 | 0.0198 | −2393 | 13.474 | 0.014 | 2456699.7581 | 0.0041 | 0.0118 | 613 | 13.358 | 0.021 |
| 2455353.4264 | 0.0028 | 0.0209 | −2302 | 13.522 | 0.022 | 2456717.7616 | 0.0023 | 0.0026 | 652 | 13.17 | 0.015 |
| 2455528.9345 | 0.0070 | 0.0198 | −1922 | 13.364 | 0.026 | 2456734.8635 | 0.0045 | 0.0154 | 689 | 13.213 | 0.024 |
| 2455534.0216 | 0.0032 | 0.0263 | −1911 | 13.281 | 0.022 | 2456737.6330 | 0.0036 | 0.0137 | 695 | 13.273 | 0.022 |
| 2455577.8959 | 0.0040 | 0.0233 | −1816 | 13.268 | 0.027 | 2456750.5609 | 0.0031 | 0.0093 | 723 | 13.373 | 0.018 |
| 2455583.8938 | 0.0038 | 0.0170 | −1803 | 13.129 | 0.032 | 2456753.3302 | 0.0060 | 0.0074 | 729 | 13.342 | 0.033 |
| 2455589.8991 | 0.0025 | 0.0180 | −1790 | 13.068 | 0.022 | 2456753.7909 | 0.0033 | 0.0063 | 730 | 13.365 | 0.019 |
| 2455607.9143 | 0.0054 | 0.0204 | −1751 | 13.286 | 0.025 | 2456754.7192 | 0.0055 | 0.0108 | 732 | 13.333 | 0.048 |
| 2455608.8384 | 0.0045 | 0.0208 | −1749 | 13.293 | 0.030 | 2456757.4857 | 0.0041 | 0.0061 | 738 | 13.258 | 0.030 |
| 2455624.5386 | 0.0030 | 0.0175 | −1715 | 13.234 | 0.027 | 2456757.4866 | 0.0033 | 0.0070 | 738 | 13.3 | 0.028 |
| 2455632.8512 | 0.0031 | 0.0165 | −1697 | 13.093 | 0.022 | 2456758.4116 | 0.0065 | 0.0083 | 740 | 13.251 | 0.043 |
| 2455637.9380 | 0.0034 | 0.0228 | −1686 | 13.152 | 0.024 | 2456763.4845 | 0.0036 | 0.0007 | 751 | 13.088 | 0.041 |
| 2455642.5675 | 0.0045 | 0.0336 | −1676 | 13.207 | 0.028 | 2456763.4866 | 0.0031 | 0.0028 | 751 | 13.152 | 0.027 |
| 2455643.4901 | 0.0024 | 0.0325 | −1674 | 13.279 | 0.011 | 2456764.4098 | 0.0024 | 0.0022 | 753 | 13.131 | 0.023 |
| 2455644.4129 | 0.0026 | 0.0316 | −1672 | 13.284 | 0.010 | 2456764.4142 | 0.0043 | 0.0066 | 753 | 13.091 | 0.038 |
| 2455645.3363 | 0.0050 | 0.0312 | −1670 | 13.288 | 0.043 | 2456767.6445 | 0.0022 | 0.0039 | 760 | 13.048 | 0.025 |
| 2455648.5686 | 0.0050 | 0.0305 | −1663 | 13.281 | 0.025 | 2456772.7275 | 0.0021 | 0.0063 | 771 | 13.054 | 0.019 |
| 2455648.5703 | 0.0060 | 0.0322 | −1663 | 13.257 | 0.037 | 2456778.7420 | 0.0033 | 0.0166 | 784 | 13.195 | 0.021 |
| 2455650.4122 | 0.0058 | 0.0266 | −1659 | 13.467 | 0.013 | 2456781.5139 | 0.0039 | 0.0173 | 790 | 13.281 | 0.018 |
| 2455656.8727 | 0.0043 | 0.0210 | −1645 | 13.288 | 0.020 | 2456782.4395 | 0.0048 | 0.0191 | 792 | 13.195 | 0.026 |
| 2455660.5618 | 0.0026 | 0.0151 | −1637 | 13.306 | 0.015 | 2456794.4377 | 0.0079 | 0.0088 | 818 | 13.305 | 0.052 |
| 2455661.4891 | 0.0038 | 0.0187 | −1635 | 13.324 | 0.014 | 2456808.7503 | 0.0026 | 0.0036 | 849 | 13.109 | 0.024 |
| 2455662.4125 | 0.0070 | 0.0184 | −1633 | 13.306 | 0.036 | 2456816.6050 | 0.0019 | 0.0065 | 866 | 13.075 | 0.018 |
| 2455668.8806 | 0.0043 | 0.0203 | −1619 | 13.205 | 0.036 | 2456828.6257 | 0.0036 | 0.0187 | 892 | 13.3 | 0.018 |
| 2455671.6500 | 0.0034 | 0.0185 | −1613 | 13.139 | 0.034 | 2457018.9021 | 0.0031 | 0.0061 | 1304 | 13.214 | 0.025 |
| 2455677.6537 | 0.0026 | 0.0180 | −1600 | 13.103 | 0.021 | 2457037.8485 | 0.0025 | 0.0160 | 1345 | 13.211 | 0.022 |
| 2455716.4519 | 0.0050 | 0.0194 | −1516 | 13.325 | 0.016 | 2457049.8734 | 0.0030 | 0.0324 | 1371 | 13.267 | 0.012 |
| 2455933.9852 | 0.0040 | 0.0136 | −1045 | 13.097 | 0.037 | 2457054.9436 | 0.0047 | 0.0220 | 1382 | 13.202 | 0.050 |
| 2455982.9391 | 0.0023 | 0.0096 | −939 | 13.126 | 0.018 | 2457074.7947 | 0.0023 | 0.0129 | 1425 | 13.207 | 0.015 |
| 2455983.8664 | 0.0033 | 0.0132 | −937 | 13.136 | 0.025 | 2457080.8020 | 0.0024 | 0.0159 | 1438 | 13.217 | 0.018 |
| 2456019.4246 | 0.0014 | 0.0077 | −860 | 13.201 | 0.012 | 2457081.7261 | 0.0031 | 0.0163 | 1440 | 13.19 | 0.038 |
| 2456254.9758 | 0.0022 | 0.0070 | −350 | 13.375 | 0.009 | 2457082.6492 | 0.0027 | 0.0156 | 1442 | 13.193 | 0.028 |
| 2456261.9102 | 0.0021 | 0.0134 | −335 | 13.467 | 0.015 | 2457091.8921 | 0.0044 | 0.0212 | 1462 | 13.243 | 0.051 |
| 2456272.9948 | 0.0010 | 0.0132 | −311 | 13.264 | 0.009 | 2457093.7481 | 0.0033 | 0.0297 | 1466 | 13.234 | 0.022 |
| 2456273.9171 | 0.0015 | 0.0117 | −309 | 13.236 | 0.009 | 2457101.5879 | 0.0028 | 0.0178 | 1483 | 13.178 | 0.025 |
| 2456279.9123 | 0.0011 | 0.0027 | −296 | 13.006 | 0.008 | 2457104.8167 | 0.0039 | 0.0135 | 1490 | 13.152 | 0.030 |
| 2456290.9961 | 0.0023 | 0.0017 | −272 | 13.117 | 0.014 | 2457106.6599 | 0.0025 | 0.0093 | 1494 | 13.133 | 0.021 |
| 2456298.8521 | 0.0022 | 0.0060 | −255 | 13.375 | 0.010 | 2457124.6781 | 0.0021 | 0.0147 | 1533 | 13.233 | 0.015 |
| 2456309.0270 | 0.0031 | 0.0198 | −233 | 13.447 | 0.010 | 2457125.6022 | 0.0020 | 0.0151 | 1535 | 13.238 | 0.015 |
| 2456309.9508 | 0.0024 | 0.0199 | −231 | 13.443 | 0.010 | 2457128.3765 | 0.0056 | 0.0182 | 1541 | 13.199 | 0.049 |
| 2456310.8747 | 0.0020 | 0.0200 | −229 | 13.423 | 0.008 | 2457129.7641 | 0.0028 | 0.0202 | 1544 | 13.195 | 0.022 |
| 2456315.0281 | 0.0017 | 0.0166 | −220 | 13.34 | 0.007 | 2457130.6878 | 0.0020 | 0.0201 | 1546 | 13.267 | 0.012 |
| 2456315.9498 | 0.0018 | 0.0146 | −218 | 13.319 | 0.009 | 2457131.6168 | 0.0034 | 0.0254 | 1548 | 13.243 | 0.023 |
| 2456323.7893 | 0.0013 | 0.0024 | −201 | 13.009 | 0.012 | 2457132.5395 | 0.0056 | 0.0244 | 1550 | 13.268 | 0.032 |
| 2456334.8733 | 0.0015 | 0.0016 | −177 | 13.105 | 0.007 | 2457133.4715 | 0.0071 | 0.0326 | 1552 | 13.259 | 0.048 |
| 2456339.0304 | 0.0019 | 0.0019 | −168 | 13.256 | 0.011 | 2457134.3925 | 0.0070 | 0.0299 | 1554 | 13.251 | 0.054 |
| 2456339.9558 | 0.0019 | 0.0035 | −166 | 13.274 | 0.009 | 2457149.6183 | 0.0023 | 0.0141 | 1587 | 13.108 | 0.021 |
| 2456340.8788 | 0.0017 | 0.0028 | −164 | 13.298 | 0.008 | 2457166.7087 | 0.0039 | 0.0154 | 1624 | 13.222 | 0.030 |
| 2456351.9774 | 0.0029 | 0.0166 | −140 | 13.46 | 0.015 | | | | | | |

Table 9. TU Com brightness maxima derived from SuperWASP database.

| Maximum HJD | Error | O–C (day) | E | Maximum HJD | Error | O–C (day) | E |
|---|---|---|---|---|---|---|---|
| 2453130.5345 | 0.0079 | 0.09241 | –7115 | 2454150.7320 | 0.0033 | 0.026854 | –4906 |
| 2453137.4585 | 0.0022 | 0.088413 | –7100 | 2454156.7298 | 0.0046 | 0.02039 | –4893 |
| 2453144.4709 | 0.0167 | 0.172816 | –7085 | 2454157.6558 | 0.005 | 0.022657 | –4891 |
| 2453174.4205 | 0.0052 | 0.101095 | –7020 | 2454158.5747 | 0.0035 | 0.017824 | –4889 |
| 2453831.5834 | 0.0069 | 0.027993 | –5597 | 2454165.5038 | 0.0079 | 0.018927 | –4874 |
| 2453832.4998 | 0.0032 | 0.02066 | –5595 | 2454169.6550 | 0.0055 | 0.013329 | –4865 |
| 2453833.4265 | 0.0035 | 0.023627 | –5593 | 2454170.5818 | 0.0067 | 0.016396 | –4863 |
| 2453856.5096 | 0.0069 | 0.013403 | –5543 | 2454171.5038 | 0.005 | 0.014663 | –4861 |
| 2454101.7852 | 0.0084 | 0.037901 | –5012 | 2454194.6109 | 0.0103 | 0.028439 | –4811 |
| 2454114.6964 | 0.0029 | 0.01684 | –4984 | 2454195.5406 | 0.0092 | 0.034406 | –4809 |
| 2454115.6165 | 0.0027 | 0.013207 | –4982 | 2454202.4539 | 0.0032 | 0.019709 | –4794 |
| 2454120.7032 | 0.0063 | 0.019376 | –4971 | 2454206.6025 | 0.0026 | 0.01151 | –4785 |
| 2454121.6234 | 0.0049 | 0.015843 | –4969 | 2454208.4512 | 0.0035 | 0.012744 | –4781 |
| 2454143.7753 | 0.0038 | –0.00185 | –4921 | 2454213.5362 | 0.0047 | 0.017213 | –4770 |
| 2454145.6606 | 0.0077 | 0.035986 | –4917 | 2454214.4571 | 0.0049 | 0.01438 | –4768 |
| 2454146.5865 | 0.0049 | 0.038153 | –4915 | 2454215.3754 | 0.0057 | 0.008947 | –4766 |